\documentclass[11pt]{article}

\usepackage[a4paper,top=3cm,bottom=3cm,left=3cm,right=3cm,marginparwidth=1.75cm]{geometry}
\usepackage[utf8]{inputenc} 
\usepackage[T1]{fontenc}    
\usepackage{hyperref}       
\usepackage{url}            
\usepackage{booktabs}       
\usepackage{amsfonts}       
\usepackage{nicefrac}       
\usepackage{microtype}      
\usepackage{graphicx}
\usepackage{amsmath}
\usepackage[capitalize,noabbrev]{cleveref}
\usepackage{microtype}
\usepackage{graphicx}
\usepackage{subcaption}
\usepackage{booktabs} 
\usepackage{multirow}
\usepackage{multicol}
\usepackage{amssymb}
\usepackage{amsthm}
\usepackage{hyperref}
\usepackage{algorithm}
\usepackage{algorithmic}
\usepackage{subcaption}
\usepackage{physics}
\usepackage{pifont}
\usepackage{xspace}
\usepackage[table]{xcolor}
\usepackage{authblk}  
\usepackage{wrapfig}
\usepackage{natbib}

\usepackage[most]{tcolorbox}
\usepackage{listings}

\lstdefinestyle{librettoGrammar}{
    basicstyle=\ttfamily\scriptsize,
    columns=fullflexible,
    keepspaces=true,
    showstringspaces=false,
    breaklines=true,
    breakatwhitespace=false,
    frame=none,
    xleftmargin=0pt,
    xrightmargin=0pt,
    aboveskip=0pt,
    belowskip=0pt
}

\newtcolorbox{grammarpanel}[2][]{
    enhanced,
    colback=gray!3,
    colframe=black!65,
    coltitle=black,
    colbacktitle=gray!18,
    fonttitle=\bfseries,
    title={#2},
    boxrule=0.6pt,
    arc=2mm,
    left=1.4mm,
    right=1.4mm,
    top=1mm,
    bottom=1mm,
    before skip=0.9em,
    after skip=1.0em,
    #1
}

\newcommand{\panelnote}[1]{
    \noindent{\small #1}
    \par\vspace{0.6em}
}

\newcommand{\panelread}[1]{
    \vspace{0.4em}
    \noindent{\small\emph{Read:} #1}
}

\newcommand{\lit}[1]{\texttt{\detokenize{#1}}}

\newcommand{\card}[1]{\left|#1\right|}
\newcommand{\indic}[1]{\mathbf{1}\!\left\{#1\right\}}
\newcommand{\qntl}[2]{Q_{#1}\!\left(#2\right)}
\newcommand{\clipthree}[3]{\min\!\left\{#3,\max\!\left\{#2,#1\right\}\right\}}
\DeclareMathOperator{\Mean}{mean}
\DeclareMathOperator{\Std}{std}
\DeclareMathOperator{\Hist}{hist}
\DeclareMathOperator{\Round}{round}
\DeclareMathOperator{\Prom}{prom}

\theoremstyle{plain}

\theoremstyle{definition}

\theoremstyle{remark}

\crefalias{lem}{thm}
\crefalias{prop}{thm}
\crefalias{cor}{thm}
\crefalias{definition}{thm}
\crefalias{asm}{thm}
\crefalias{rem}{thm}

\crefname{thm}{theorem}{theorems}
\Crefname{thm}{Theorem}{Theorems}

\crefname{prop}{proposition}{propositions}
\Crefname{prop}{Proposition}{Propositions}

\crefname{lem}{lemma}{lemmas}
\Crefname{lem}{Lemma}{Lemmas}

\crefname{cor}{corollary}{corollaries}
\Crefname{cor}{Corollary}{Corollaries}

\crefname{definition}{definition}{definitions}
\Crefname{definition}{Definition}{Definitions}

\crefname{asm}{assumption}{assumptions}
\Crefname{asm}{Assumption}{Assumptions}

\crefname{rem}{remark}{remarks}
\Crefname{rem}{Remark}{Remarks}

\title{Libretto: Giving LLM Agents a Sense of Musical Structure}

\makeatletter
\patchcmd{\AB@output}{\par\vspace{1em}}{}{}{}
\renewcommand\AB@affilsepx{\quad}  
\makeatother

\author[1]{Yichen Xu}
\affil[1]{University of California, Berkeley}

\begin{document}

\date{}
\maketitle

\begin{abstract}
Generative music systems can now produce impressive audio from text prompts, but audio outputs are difficult to inspect, edit, and diagnose as musical structure. We introduce \textsc{Libretto}, an agent-facing framework for symbolic music generation and revision. Libretto uses an LLM-native grammar with explicit onset slots, voices, and bar-level organization, then evaluates each piece in a corpus-calibrated statistical space over rhythm, harmony, melody, texture, form, and variation. The same structural axes support retrieval, diagnosis, copy-risk control, and iterative self-revision. Across gap filling, reference-guided full-piece generation, gradual morphing, and educational music generation, Libretto turns symbolic music from a raw token sequence into a measurable and editable object for language-model agents.
\end{abstract}


\section{Introduction}

Generative AI has made music creation increasingly accessible. Commercial systems such as Suno, Udio, Stable Audio, Eleven Music, and MusicFX can produce complete audio from short prompts, and research systems such as MusicLM and MusicGen show strong text-conditioned audio generation capabilities \citep{suno,udio,stableaudio,elevenmusic,musicfx,agostinelli23,copet23}. These systems are powerful, but their audio outputs are difficult to inspect as musical objects. A waveform can be heard and rated, but it does not directly expose note timing, voice assignment, phrase structure, harmonic motion, repetition, or local edit boundaries.

Symbolic music offers a complementary path because it keeps music in an editable representation. Prior work has generated symbolic music with chorale models, multi-track GANs, latent musical spaces, long-range Transformers, beat-aware event formats, and multi-track orchestral representations \citep{had17,dong18,pmlr-v80-roberts18a,huang18,huang20,yu22,liu22}. More recent work connects symbolic music with large language models through ABC-based music-language models, symbolic pretraining, text-to-MIDI adaptation, and multi-agent composition \citep{yuan24,qu25,wu25,deng24,xing25}. However, most trained sequence models require specialized deployment, and current agentic systems leave open a basic interface question: what symbolic form should an agent read, write, and revise? Compact formats such as ABC are useful, but their timing can be implicit, making onset reasoning and local editing harder for a language-model agent.

Evaluation is another limitation. Existing systems often rely on human preference, likelihood, prompt adherence, contrastive audio-text alignment, or learned aesthetic predictors \citep{elizalde23,tjandra25}. These measurements are useful, but they do not always explain what structural property of a generated piece failed or how an agent should revise it. Music theory has long treated music as organized structure across meter, grouping, harmony, and repetition \citep{lerdahl83}. This motivates an evaluation interface that describes music through interpretable structural properties rather than only through a global quality score.

\begin{figure}[t]
    \centering
    \includegraphics[width=\linewidth]{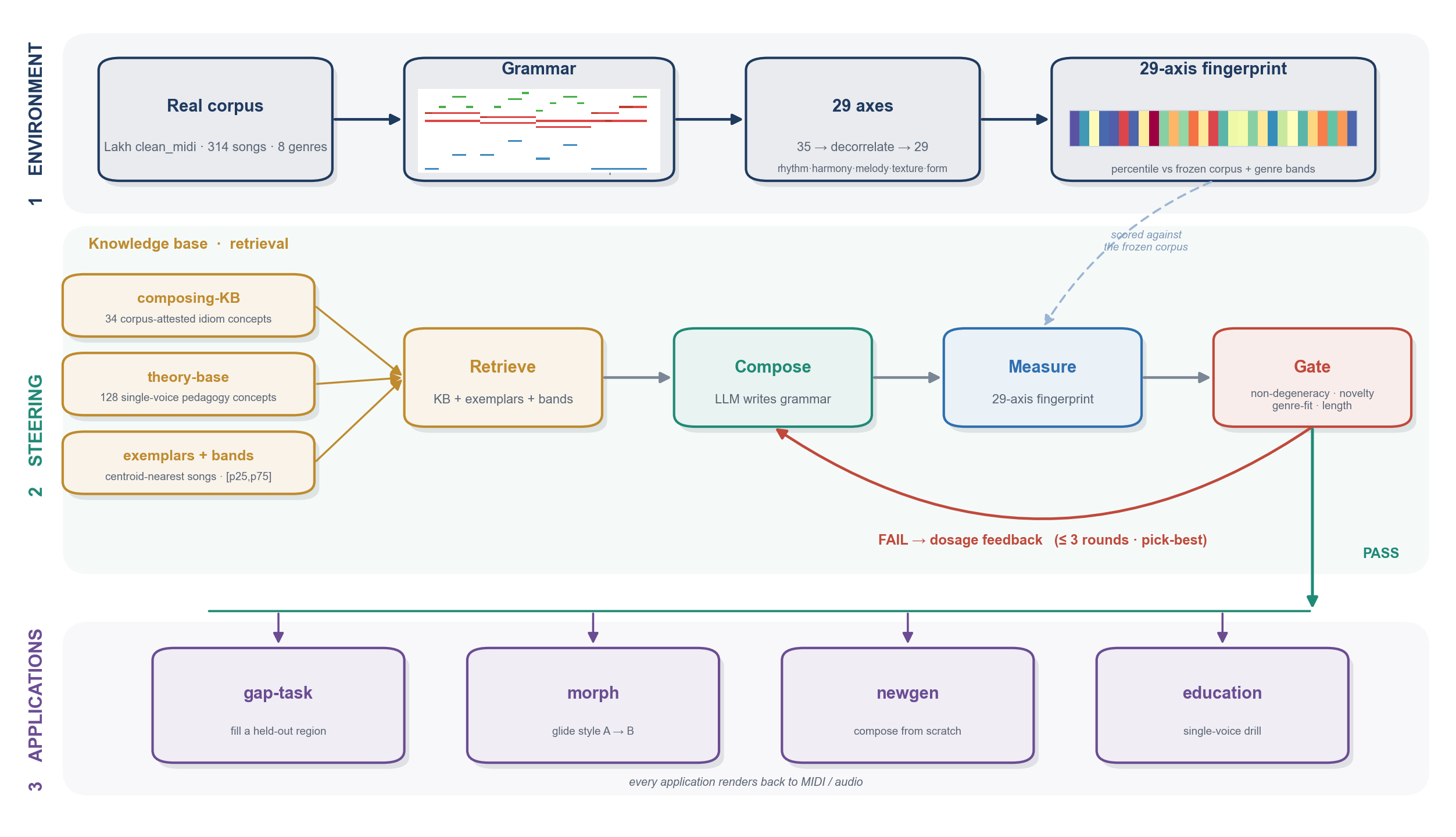}
    \caption{Overview of the \textsc{Libretto} workflow.}
    \label{fig:workflow}
\end{figure}

We introduce \textsc{Libretto}, an agent-facing framework for symbolic music generation and revision. Libretto represents each piece in an LLM-native grammar with explicit onset slots, voices, and bar-level organization, so that timing and structure are directly readable and locally editable. It then places each piece in a corpus-calibrated statistical cloud: a set of interpretable axes over rhythm, harmony, melody, texture, form, and within-song variation. These axes make music structure measurable, allowing generation to be diagnosed by where it falls relative to existing music. \Cref{fig:workflow} gives an overview of the full workflow, from symbolic representation and retrieval to measurement, feedback, and revision.

Our contributions are threefold:
\begin{itemize}
    \item We characterize symbolic music through a corpus-calibrated statistical cloud rather than a subjective quality score. Each piece is located along interpretable axes of rhythm, harmony, melody, texture, form, and variation, enabling generation to be diagnosed by where it falls relative to existing music. An LLM-native grammar serves as the interface through which these axes become readable and editable by a language-model agent, while also supporting straightforward downstream MIDI rendering.

    \item We build an agentic composition system that combines knowledge bases, retrieval, measurement, and iterative self-improvement. The agent retrieves relevant musical concepts and examples, generates a candidate, evaluates it using the structural axes, and revises it through musician-readable feedback.

    \item We show that the framework supports multiple symbolic-music applications, including gap filling, reference-guided new-piece generation, gradual morphing between musical styles or pieces, and educational music generation for targeted theory concepts. The same framework also supports longer-form generation than the short fixed-length settings common in prior work, with new multi-voice pieces typically spanning around 100 bars.
\end{itemize}

\section{Related Work}

\paragraph{Symbolic music generation and representations.}
Deep learning for symbolic music generation has been studied through many representations, model families, and evaluation protocols; we refer readers to the survey by \citet{ji23} for a broader overview. Early Transformer-based work such as Music Transformer showed that relative self-attention can generate symbolic music with long-term structure, motif continuation, and coherent accompaniment \citep{huang18}. Subsequent work emphasized that representation design is as important as model architecture. Pop Music Transformer introduced REMI, a beat-based event representation with explicit bar, position, tempo, and chord tokens, making metrical and harmonic structure easier for a sequence model to learn \citep{huang20}. Museformer addressed long symbolic sequences by combining fine-grained attention over structure-related bars with coarse-grained summaries of other bars \citep{yu22}. SymphonyNet focused on complex orchestral scores, proposing a multi-track, multi-instrument representation and permutation-aware modeling for symphonic generation \citep{liu22}. Anticipatory Music Transformer treated symbolic music as a temporal point process and used arrival-time tokenization for controllable infilling, making onset time explicit in the event sequence \citep{thickstun24}. MuPT scaled ABC-based symbolic pretraining and introduced synchronized multi-track ABC to improve bar alignment across tracks \citep{qu25}. Libretto follows this line of work in treating representation as central, but it targets a different interface: the grammar is designed for language-model agents to read, edit, and diagnose musical structure directly.

\paragraph{LLMs and agentic symbolic composition.}
Recent work has explored whether large language models can understand and generate symbolic music as text. ChatMusician continues pretraining and fine-tuning LLaMA2 on ABC notation and music-language data, treating music as a second language and evaluating both generation and music-theory understanding \citep{yuan24}. NotaGen adopts large-language-model training paradigms for symbolic generation, including pretraining and finetuning to improve musicality \citep{wang25}. MIDI-LLM instead expands a pretrained text LLM with MIDI tokens and trains it for text-to-MIDI generation, using explicit onset, duration, and instrument-pitch tokens \citep{wu25}. MetaScore builds a large symbolic-score dataset with rich metadata and LLM-generated captions, then trains text- and tag-conditioned symbolic music generators \citep{xu25}. In parallel, ComposerX and CoComposer explore training-free multi-agent composition: ComposerX decomposes symbolic composition into leader, melody, harmony, instrument, review, and arrangement agents \citep{deng24}, while CoComposer uses leader, melody, accompaniment, revision, and review agents and evaluates generated outputs with AudioBox-Aesthetics \citep{xing25}. Libretto is closest to these LLM-based symbolic systems, but shifts the contribution from model training or agent role design to the representation-evaluation loop: generated music is written in a directly inspectable grammar, measured by corpus-calibrated structural axes, and revised through explicit feedback.

\paragraph{Text-to-audio generation, benchmarks, and evaluation.}
Text-conditioned audio generators provide another important comparison point. MusicGen models compressed audio tokens with a single-stage Transformer and supports text and melody conditioning, producing strong audio outputs but not directly exposing symbolic structure for editing or diagnosis \citep{copet23}. Automatic audio evaluation has also developed in this direction: AudioBox-Aesthetics predicts production quality, production complexity, content enjoyment, and content usefulness for speech, music, and sound, offering a scalable alternative to human listening tests \citep{tjandra25}. For symbolic-music understanding, ABC-Eval benchmarks LLMs on ABC notation across syntax, segment-level, and sequence-level tasks, showing that text-based symbolic music reasoning remains difficult for current models \citep{zhao25}. These works motivate Libretto's focus on symbolic structure rather than only audio quality or prompt adherence. Libretto does not replace audio-domain systems or trained symbolic generators; instead, it provides a text interface where timing, voices, and structural measurements remain available throughout retrieval, generation, diagnosis, and self-revision.

\section{Methods}
\label{sec:method}

Libretto is a text interface for symbolic music. It converts MIDI-like note structure into an LLM-readable grammar, places each piece inside a corpus-calibrated statistical music cloud, and uses that position to guide an agent through retrieval, generation, diagnosis, and revision. The method consists of five parts: the grammar, the structural axis system, the agent loop, the knowledge bases, and the application-specific task setups. We use Claude Code with Opus 4.8 as the LLM agent throughout all experiments, and use 314 real MIDI files spanning eight genres as the raw music corpus, curated from the Lakh MIDI Dataset \citep{Raffel16}. 

\paragraph{Grammar.}
Libretto represents a piece as plain text with a global header, a voice declaration, and one block per bar. The header specifies key, meter, tempo, grid, and bar count; the voice line declares the ordered parts; and each bar contains a required chord label followed by voice-specific note tokens. Each token specifies pitch, onset slot, and duration slot, while simultaneous pitches are joined with a plus sign. The grammar uses integer slots rather than floating-point time. In a 16th-note 4/4 grid, for example, the beat positions are slots 1, 5, 9, and 13. This makes rhythm compact and explicit, while preserving separate voices for bass lines, melody, accompaniment, and other parts. The encoder also supports adaptive grids, including triplet grids, so that timing can be preserved without making the text unnecessarily dense. The representation is faithful to pitch, quantized onset, quantized duration, and voice separation, and deliberately abstracts away velocity, micro-timing, original timbre, and unpitched percussion. Its goal is to expose score-like structure in a form that an LLM agent can read, edit, and regenerate. Example grammar panels are shown in appendix \Cref{box:education-drill,box:new-generation,box:gap-fill}.

\paragraph{Structural axes.}
Libretto evaluates music by locating it inside a statistical cloud of existing pieces rather than assigning a subjective quality score. Each piece is mapped to a 29-axis fingerprint covering rhythm, harmony, melody, texture, form, and within-song variation. These axes are computed directly from grammar tokens, so they describe observable musical structure rather than human preference. Candidate axes are selected by two principles: they must vary meaningfully across the corpus, and they should not be redundant with one another. Near-constant metrics are removed, and highly correlated metrics are pruned, leaving a compact set of relatively independent structural descriptors. Each raw axis value is then converted to a percentile against a frozen 314-song corpus, giving all axes a common distribution-free scale. A percentile is descriptive: high or low does not mean good or bad. Extreme percentiles are used only to diagnose structural degeneracy, such as outputs that become unusually repetitive, dense, sparse, harmonically unstable, or rhythmically atypical relative to real music. The metric calculation can be found in \Cref{app:metrics}.

\paragraph{Agent loop.}
Generation is organized as a bounded generate--measure--revise loop. The agent first produces a candidate grammar; the system parses it, computes the structural fingerprint, checks copy risk, and evaluates task-specific gates; then the next prompt receives musician-readable feedback. The feedback does not expose raw metric identifiers or ask the model to hit exact numerical targets. Instead, it describes musical tendencies to adjust, such as making the texture less sparse, reducing harmonic instability, increasing genre fit, or moving an extreme axis back toward the idiomatic middle. The loop keeps the best candidate found so far, so refinement is safe at selection time: a later attempt is retained only if it improves the measured structural score.

\paragraph{Knowledge bases and retrieval.}
Retrieval gives the agent concrete musical grounding. Libretto uses two knowledge bases. The composing knowledge base is corpus-grounded and contains idiomatic concepts for harmony, groove, melody, voicing, form, and jazz-specific techniques. Each entry includes corpus attestation, real examples, and an actionable composition instruction; it is used for genre-conditioned generation and morphing. The theory knowledge base is pedagogical and contains single-voice examples for scales, chords, progressions, rhythm patterns, cadences, texture, and form. Each entry includes a challenge that the generated drill must satisfy; it is used for education tasks. For new-piece generation, Libretto also retrieves short real excerpts from songs nearest to the target genre centroid in fingerprint space. These excerpts provide style references, while copy-risk gates prevent direct reuse.

\paragraph{Copy and novelty.}
Libretto separates idiomatic similarity from direct copying. Copy risk is measured at the note level by comparing bar-aligned onset--pitch pairs between generated material and real references. The system checks overlap against retrieved examples, likely corpus matches, and, in gap filling, the hidden answer region. This is stricter than comparing chord labels or bar-level summaries: a piece may share genre idioms, harmonic language, or rhythmic feel, but it should not reproduce the same note-level material. In education, the same principle is applied against the shown theory example, so the drill must instantiate the requested concept without simply copying the demonstration.

\paragraph{Applications.}
The same grammar, statistical axes, retrieval mechanism, and refinement loop support four tasks. In gap filling, the model receives surrounding musical context and fills a held-out region. The output must match the missing length, fit the local context, avoid structural degeneracy, and avoid copying either the hidden answer or corpus material. In new generation, the model composes a full piece in a target genre from scratch, using retrieved concepts and prototypical excerpts as references; the loop pushes the result away from extremes, weak genre fit, and copy risk. In our experiments, new-generation outputs typically span 92--128 bars, with a mean length of about 102 bars. In morphing, the model gradually moves from one source style or component toward another, and evaluation checks source-like beginning, target-like ending, smooth progress, and non-abrupt transition. In education, the model generates a short drill for a requested theory concept, requiring valid grammar, key adherence, satisfaction of user constraints, detection of the target concept, and novelty relative to the shown example.

\section{Experiments}

\subsection{Representation and evaluation}

Before evaluating generation, we first validate the text grammar itself. The audit tests whether the representation preserves the musical information used by the downstream language-model pipeline, including pitch, timing, and voice assignment, while also making explicit which musical dimensions are intentionally abstracted away. We also test timing readability. In ABC, note starts must be recovered by accumulating earlier durations in the bar, while in Libretto each note directly carries its absolute onset slot. We then define a 29-dimensional structural fingerprint, where each axis is evaluated as a corpus percentile against a fixed 314-song reference set. Values near 50 are corpus-typical, while values at or below the 5th percentile or at or above the 95th percentile are treated as structural extremes.

\begin{table}[t]
\centering
\footnotesize
\renewcommand{\arraystretch}{1.06}
\caption{Representation and metric validation. The grammar preserves downstream musical structure with quantified losses, makes timing directly readable, and the 29-axis fingerprint is low-redundancy but genre-informative.}
\label{tab:representation_metric}
\resizebox{\linewidth}{!}{
\begin{tabular}{p{0.21\linewidth}p{0.30\linewidth}p{0.41\linewidth}}
\toprule
Check & Validation protocol & Result \\
\midrule
Pitch and voice
& 18 files; 9 genres; 57{,}885 non-drum notes
& Pitch set preserved exactly; 11 notes lost, or 0.019\%; source parts map 1:1 to grammar voices. \\
\midrule
Timing fidelity
& Round-trip note starts and durations
& Start-time error after round trip: median 0 ms, mean 3.1 ms, worst note within one grid slot; durations rounded to the bar grid. \\
\midrule
Timing readability
& ABC vs. Libretto on 5 tunes, 98 notes, 28 bars
& Recovering note starts takes 70 running additions in ABC, or 131 additions from scratch; Libretto takes 0 because onset slots are explicit. \\
\midrule
Edit locality
& Single duration edits under relative timing
& ABC causes 1005 downstream onset re-derivations over the corpus; Libretto causes 0. Cost: ABC 335 body characters, Libretto 1176. \\
\midrule
Intentional abstractions
& Dimensions not represented
& Velocity flattened; micro-timing quantized; drums dropped, mean 21\% of source notes; instrument identity reassigned. \\
\midrule
Axis construction
& Spread and redundancy filters
& 35 candidate measurements reduced to 29 axes over rhythm, harmony, melody, texture, form, and within-song variation. \\
\midrule
Axis independence
& Pairwise correlations on 314 songs
& Mean $|r|=0.16$; 12/406 retained axis pairs exceed 0.5; largest remaining $|r|=0.77$. \\
\midrule
Genre signal
& Five-fold classification on 255 songs
& Logistic regression on the 29 axes reaches 0.384 top-1 accuracy over 8 genres, about 3.1$\times$ chance. \\
\bottomrule
\end{tabular}
}
\end{table}

\Cref{tab:representation_metric} establishes the measurement substrate used in the rest of the experiments. The grammar is effectively exact for pitch and voice, grid-faithful for timing, and explicit about the musical information it abstracts away. It also makes timing local: Libretto uses more characters than ABC, but avoids the duration accumulation needed to recover note starts and prevents duration edits from shifting downstream onsets. This is therefore an encoding-cost result rather than an LLM benchmark. The structural axes are validated separately: they are filtered for spread and redundancy, then evaluated for independence and musical signal.

\begin{figure}[t]
\centering
\includegraphics[width=\linewidth]{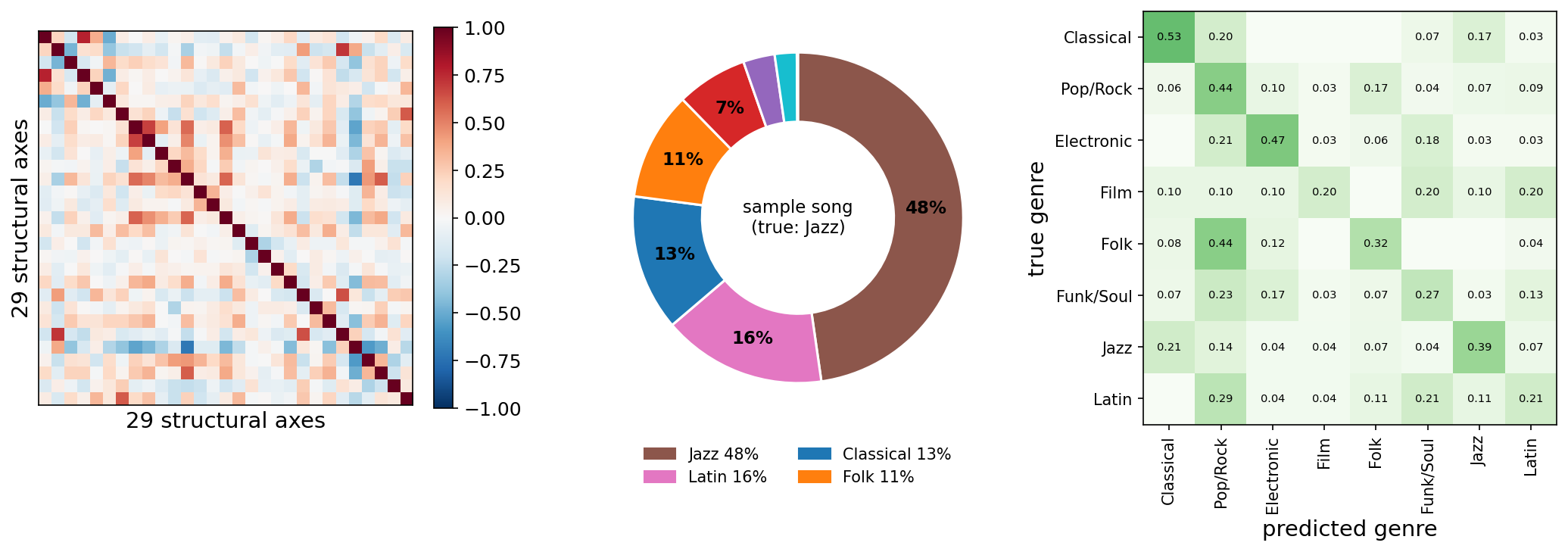}
\caption{Axis structure and soft genre signal.}
\label{fig:axis_structure}
\end{figure}

\Cref{fig:axis_structure} visualizes the 29-axis fingerprint as a measurement system. The correlation panel shows that most axis pairs are weakly related. The genre-composition panel shows that a song is represented as a soft stylistic mixture rather than as a hard category. The confusion matrix makes the same point at corpus scale: the diagonal is visible, but neighboring genres blur into one another. This behavior is desirable for a structural metric: it captures genre tendencies without forcing musical style into clean clusters.

\begin{figure}[t]
\centering
\includegraphics[width=\linewidth]{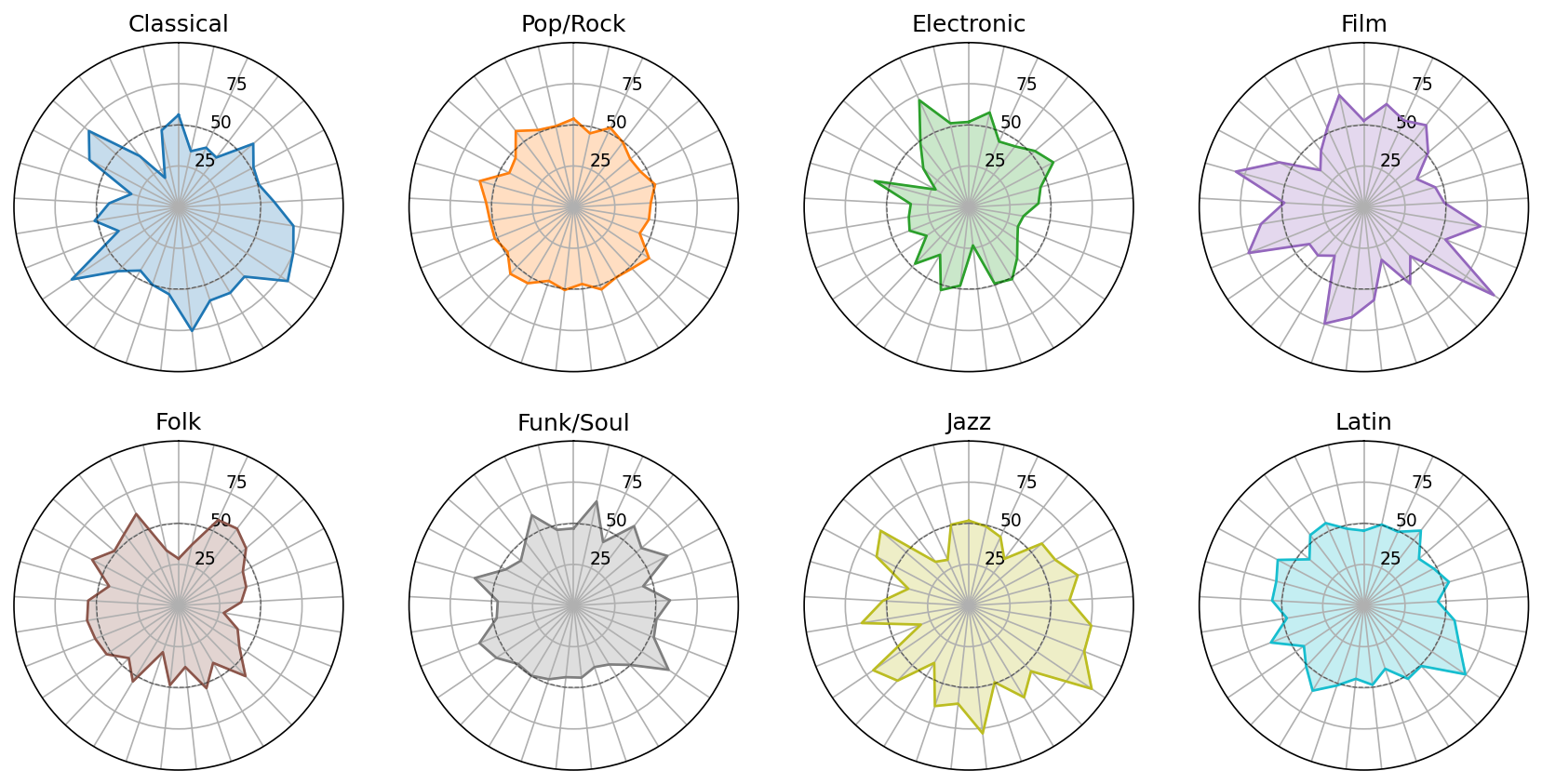}
\caption{Faceted genre fingerprints over the 29 axes.}
\label{fig:genre_radar_faceted}
\end{figure}

\Cref{fig:genre_radar_faceted} makes the axes musically interpretable. Each spoke is one structural measurement, and each value is the genre mean corpus percentile. Jazz expands on harmonic-complexity axes, with chromaticism near the 75th percentile and diminished/augmented color near the 76th percentile. Folk is lower on the same axes, around the 28th and 39th percentiles. Electronic music is highly self-similar, around the 72nd percentile, while classical is much lower, around the 20th percentile. These gaps of roughly 30--50 percentile points show that the fingerprint carries readable stylistic structure.

The pass gates use the same percentile language. The main structural gate counts how many axes land in the extreme tails of the real-song distribution. Since distinctive real music can naturally contain some extreme measurements, the budgets are calibrated against real songs rather than imposed uniformly. A separate copy-risk gate rejects outputs that are too close to the source, context, or hidden answer.

\begin{table}[t]
\centering
\footnotesize
\renewcommand{\arraystretch}{1.06}
\caption{Gate calibration and discrimination. Each gate is calibrated on real or acceptable music and targets a distinct generated-output failure mode.}
\label{tab:gates}
\resizebox{\linewidth}{!}{
\begin{tabular}{p{0.18\linewidth}p{0.38\linewidth}p{0.36\linewidth}}
\toprule
Gate & Calibration & Generated-output behavior \\
\midrule
Degeneracy
& Real songs: mean 3.4 extreme axes out of 29, median 3, 90th percentile 7; final genre budgets allow 4--6 extremes.
& Gap-task failures average 4.36 extremes versus 2.25 for passes; ungrounded full-piece generations average 5.9. \\
\midrule
Genre fit
& A fixed six-of-eight fit rule admits only 10--29\% of real songs in their own genres; final floors are genre-calibrated.
& Calibrated floors avoid rejecting human music while still requiring generated pieces to occupy genre-typical structural bands. \\
\midrule
Copy risk
& Real-song copy statistics have 90th percentile at most 0.29 across genres, grounding a threshold near 0.30.
& Passing fills: mean 0.233, max 0.294; flagged fills: mean 0.383, min 0.312. \\
\midrule
Novelty
& Accepted education drills have copy-vs-shown mean 0.065 and max 0.127, below the 0.50 novelty threshold.
& Rejects drills that reuse the shown material instead of producing a new example of the target concept. \\
\bottomrule
\end{tabular}
}
\end{table}

\Cref{tab:gates} shows that the gates are calibrated against real or acceptable music before being applied to generated outputs. Real music naturally contains some extreme axes, so the degeneracy gate allows a small calibrated budget rather than requiring every axis to be central. Genre fit is likewise calibrated per genre because a fixed rule would reject too much human music. Copy risk is anchored to real-song overlap statistics, and novelty is checked against the examples shown in the education setting. Together, these gates target the intended generated-output failures: structural collapse, weak genre fit, replication of corpus or held-out material, and copying from demonstrations.

\subsection{Application-level results}

We evaluate the grammar and gates across four applications: filling missing regions, generating complete pieces, morphing between styles, and creating education drills. A generation passes only if it satisfies the relevant structural, fit, copy, and task-specific checks. The same metric family is used across tasks, but each task stresses a different failure mode: local coherence for gap filling, non-degeneracy for full-piece generation, continuity for morphing, and requirement satisfaction for drills.

\begin{table}[t]
\centering
\footnotesize
\renewcommand{\arraystretch}{1.06}
\caption{Main application results. Retrieval and the self-evolving loop improve the settings where generation otherwise collapses or underspecifies structure.}
\label{tab:application_results}
\resizebox{\linewidth}{!}{
\begin{tabular}{p{0.22\linewidth}p{0.31\linewidth}p{0.18\linewidth}p{0.21\linewidth}}
\toprule
Application & Comparison & Baseline & Result \\
\midrule
Gap filling
& Single-shot vs. loop, up to 3 rounds
& 6/51 pass, 12\%
& 20/51 pass, 39\%; 33/51 improve. \\
\midrule
Full-piece generation
& Single-shot vs. loop, up to 3 rounds
& 10/16 pass, 62\%
& 15/16 pass, 94\%; loop runs only on failures. \\
\midrule
Full-piece generation
& Retrieval off vs. retrieval on
& 2/8 pass, 25\%
& 6/8 pass, 75\%; copy risk below gate in every genre. \\
\midrule
Education drills
& Retrieval off vs. retrieval on
& 7/8 pass
& 6/8 pass; retrieval is nearly null because concepts are already explicit. \\
\midrule
Education drills
& Retrieval-on model comparison
& All models output valid grammar
& Opus 6/8, Sonnet 3/8, Haiku 3/8 pass. \\
\midrule
Gradual morphing
& Source-to-target component morph
& --
& 11/21 pass all morph checks. \\
\bottomrule
\end{tabular}
}
\end{table}

\Cref{tab:application_results} gives the main outcomes. The loop is most useful when a first draft is close but structurally flawed: it raises the gap-filling pass rate from 12\% to 39\% and the full-piece pass rate from 62\% to 94\%. Retrieval addresses a different weakness. In full-piece generation, it triples the pass rate from 25\% to 75\% by grounding the model in concrete musical examples. In education, where the desired scale, rhythm, and challenge constraints are already explicit, retrieval adds little.

\Cref{fig:app_gallery} shows the breadth of the system in a single visual language. The examples include a passing jazz gap-fill with one extreme axis, a 96-bar jazz full-piece generation with low copy risk across rounds, an electronic-to-folk morph, and an E harmonic minor education drill with 2.2\% out-of-scale notes and copy-vs-shown score 0.065. Together, these examples show that the same text representation supports coherent structure across different musical objectives.

\begin{figure}[t]
\centering
\includegraphics[width=\linewidth]{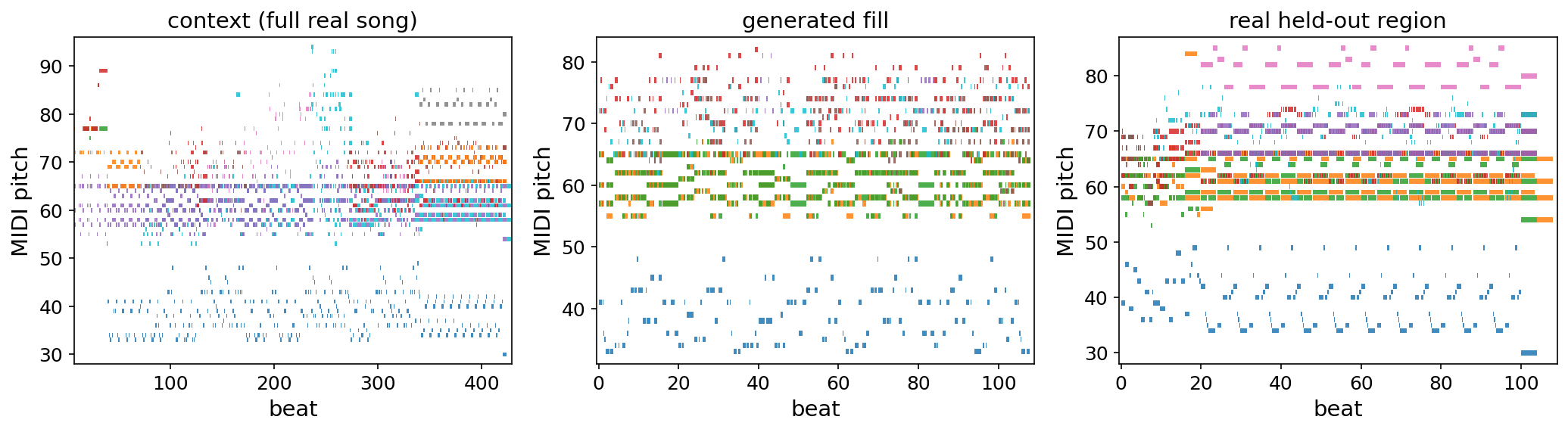}
\caption{Gap-task triptych: context, generated fill, and held-out answer.}
\label{fig:r3_gap_triptych}
\end{figure}

\Cref{fig:r3_gap_triptych} focuses on the gap-filling setting. The left panel is the surrounding real context, the middle panel is the generated fill, and the right panel is the held-out answer. The task is to land in the same musical neighborhood as the context without copying the hidden answer. This jazz continuation passes at round 3 with one extreme axis, copy risk 0.251, answer overlap 0.145, and beat alignment 98\%, showing that it fits the local texture while remaining distinct from the ground truth.

\begin{figure}[t]
\centering
\includegraphics[width=\linewidth]{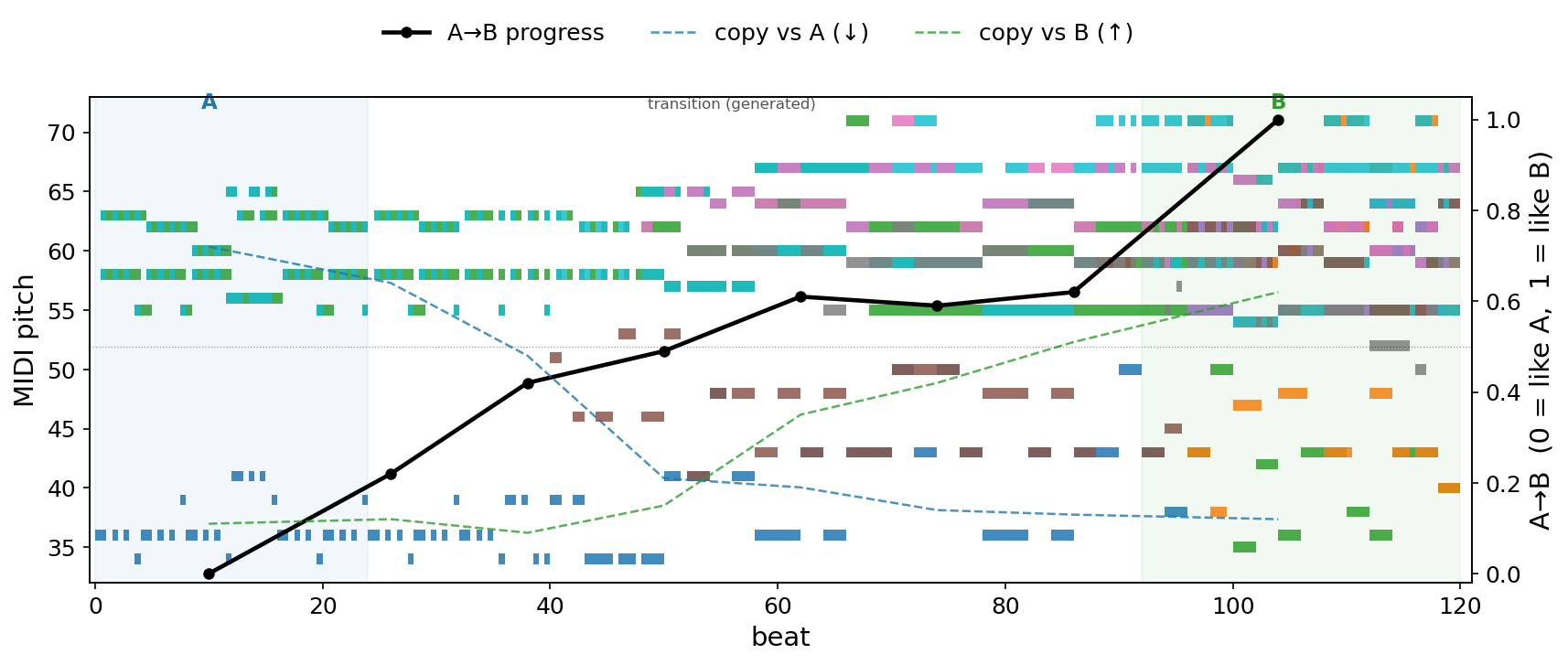}
\caption{Gradual morph with measured progress curve.}
\label{fig:r4_morph}
\end{figure}

\Cref{fig:r4_morph} illustrates the morph task, where the output should begin close to source style A, end close to target style B, and transition smoothly between them. In this electronic-to-folk example, the black A$\to$B progress curve rises across the generated region, indicating a gradual movement toward the target style. At the same time, similarity to A decreases while similarity to B increases, as shown by the two dashed copy-risk curves. The transition does not occur as a single abrupt jump; instead, the generated region steadily moves from the source-like opening toward the target-like ending.

\Cref{fig:education_examples} shows example scores generated for education drills. A student can request new practice material in different keys, modes, rhythmic patterns, or melodic settings, so the same theory concept can be practiced through fresh short scores rather than repeated from a fixed exercise.

\subsection{Mechanisms and diagnostics}

The aggregate gains in \Cref{tab:application_results} come from two different mechanisms. Retrieval helps before generation by grounding the prompt in real musical targets; the loop helps after generation by iteratively repairing outputs that fail the gates. We visualize both mechanisms at the level of the structural axes.

\Cref{fig:retrieval_mechanism} shows that retrieval improves full-piece generation by pulling extreme axes back toward the corpus band. Without retrieval, Pop/Rock collapses to the 5th percentile for mean note duration and the 95th percentile for diminished/augmented color; Film reaches the 98th percentile for ascending-step ratio, the 97th percentile for root-motion variety, and the 100th percentile for novelty. With retrieval, Pop/Rock moves to 52, 60, and 46 on the corresponding axes, while Film moves root motion from 97 to 46, ascending-step ratio from 98 to 12, maximum chord width from 1 to 59, and novelty from 100 to 74. Retrieval therefore improves pass rate by de-degenerating the fingerprint, not by increasing similarity to the corpus through copying.

\begin{figure}[t!]
    \centering
    \includegraphics[width=0.75\linewidth]{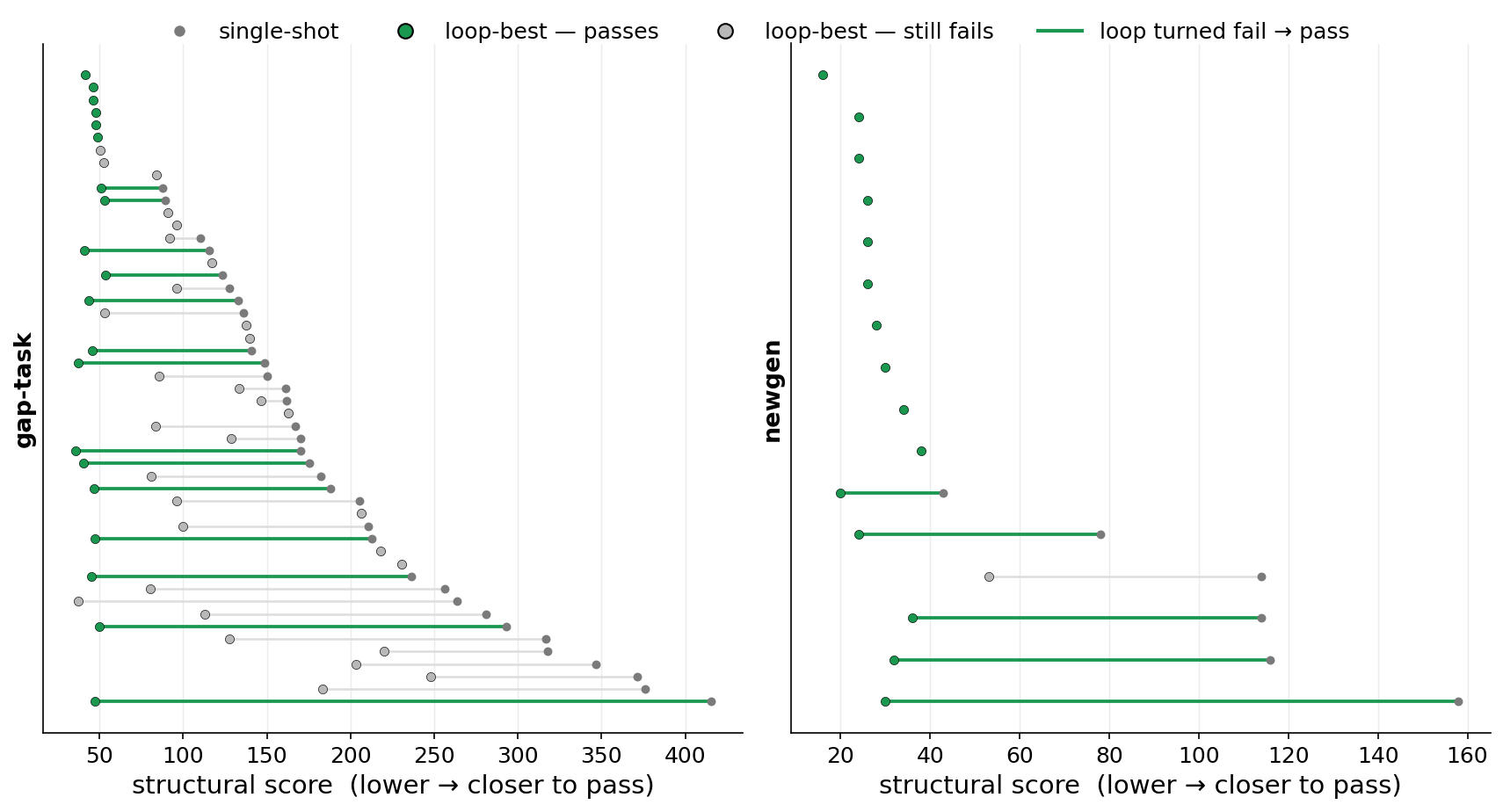}
    \caption{Per-song effect of the self-evolving loop.}
    \label{fig:loop_per_song}
\end{figure}

\Cref{fig:loop_per_song} shows the loop at the level of individual songs. Each row compares the single-shot score with the best loop candidate, and lower is better. In the gap task, 33 of 51 songs improve and the pass count rises from 6 to 20. In full-piece generation, only six rows move because the other ten already passed single-shot; among those looped pieces, five cross the pass gate. This explains why the loop is useful but not indiscriminate: it acts only where the gates expose a concrete failure.

\Cref{fig:axis_profile_heatmap} shows the same supervision signal directly. Rows are generated pieces, columns are axes, colors are corpus percentiles, and dots mark tail extremes. In the eight retrieval-on full-piece generations, the average piece has 4.6 extreme axes, but the distribution is diagnostic: electronic has one extreme, Latin and Funk/Soul have three, Pop/Rock and Film have four, and Folk accumulates 10 extremes and fails. The heatmap turns a pass/fail decision into a readable map of which musical dimensions still need repair.

\Cref{fig:r2_roll} shows why the diagnostic is musically meaningful rather than just numerical. This Latin gap-fill fails with six extreme axes against a budget of three: harmonic rhythm at the 100th percentile, chord variety at the 99th, average note length at the 96th, stepwise motion at the 95th, bar distinctness at the 100th, and density variation at the 3rd percentile. Those numbers correspond to what is visible in the roll: a chord-conveyor-belt texture with roughly two chords per bar and a scalar, highly stepwise melody. The case is not a copy, with copy risk 0.134 and answer overlap 0.098, and it is not simply off-grid, with beat alignment 86\%. The failure is structural degeneracy, and the named axes make that failure inspectable.

\section{Conclusion}

We presented \textsc{Libretto}, an agent-facing framework that makes symbolic music readable, measurable, and revisable by an LLM agent. Instead of treating generation quality as a single subjective score, Libretto represents each piece in a corpus-calibrated structural space over rhythm, harmony, melody, texture, form, and variation. This lets the agent diagnose where a candidate departs from real music, retrieve relevant examples, and revise through musician-readable feedback. Across gap filling, full-piece generation, gradual morphing, and educational music generation, the same representation-evaluation loop supports multiple symbolic-music tasks without training a new music model.

Several directions remain open. The structural axes can be refined, expanded, and pruned using the same principles used here: each added axis should capture meaningful musical variation while remaining sufficiently decorrelated from existing measurements. The feedback loop also suggests a natural path toward agentic reinforcement learning for music generation, where actions such as retrieval, editing, rewriting, and accepting can be optimized against structural rewards, copy-risk constraints, and human preference. More broadly, Libretto points toward symbolic-music agents that do not merely emit notes, but learn to reason over measurable musical structure.

\section*{Code and Website}

The project website is available at \url{https://libretto.site/}.
The code is available at \url{https://github.com/Xyc-arch/Libretto}.

\bibliographystyle{plainnat}
\bibliography{ref}

\clearpage 
\newpage
\appendix
\setcounter{figure}{0}
\renewcommand{\thefigure}{A.\arabic{figure}}

\setcounter{table}{0}
\renewcommand{\thetable}{A.\arabic{table}}
\section*{Appendix}
\label{sec:app}

\section{Auxiliary results}

\begin{figure}[t]
    \centering
    \includegraphics[width=\linewidth]{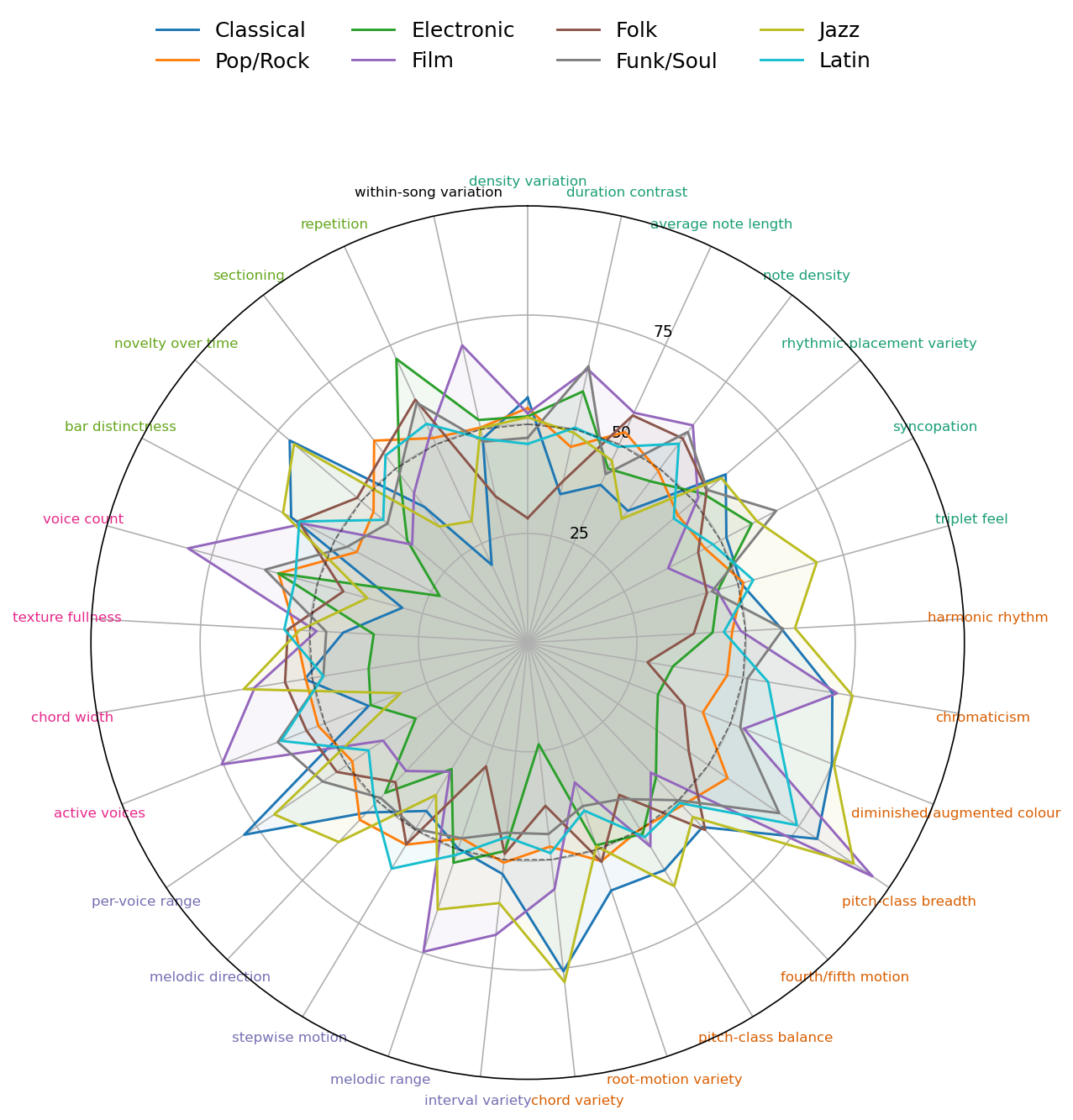}
    \caption{Overlaid genre fingerprints over the 29 axes.}
    \label{fig:genre_radar_overlaid}
\end{figure}

\begin{figure}[t]
    \centering
    \includegraphics[width=\linewidth]{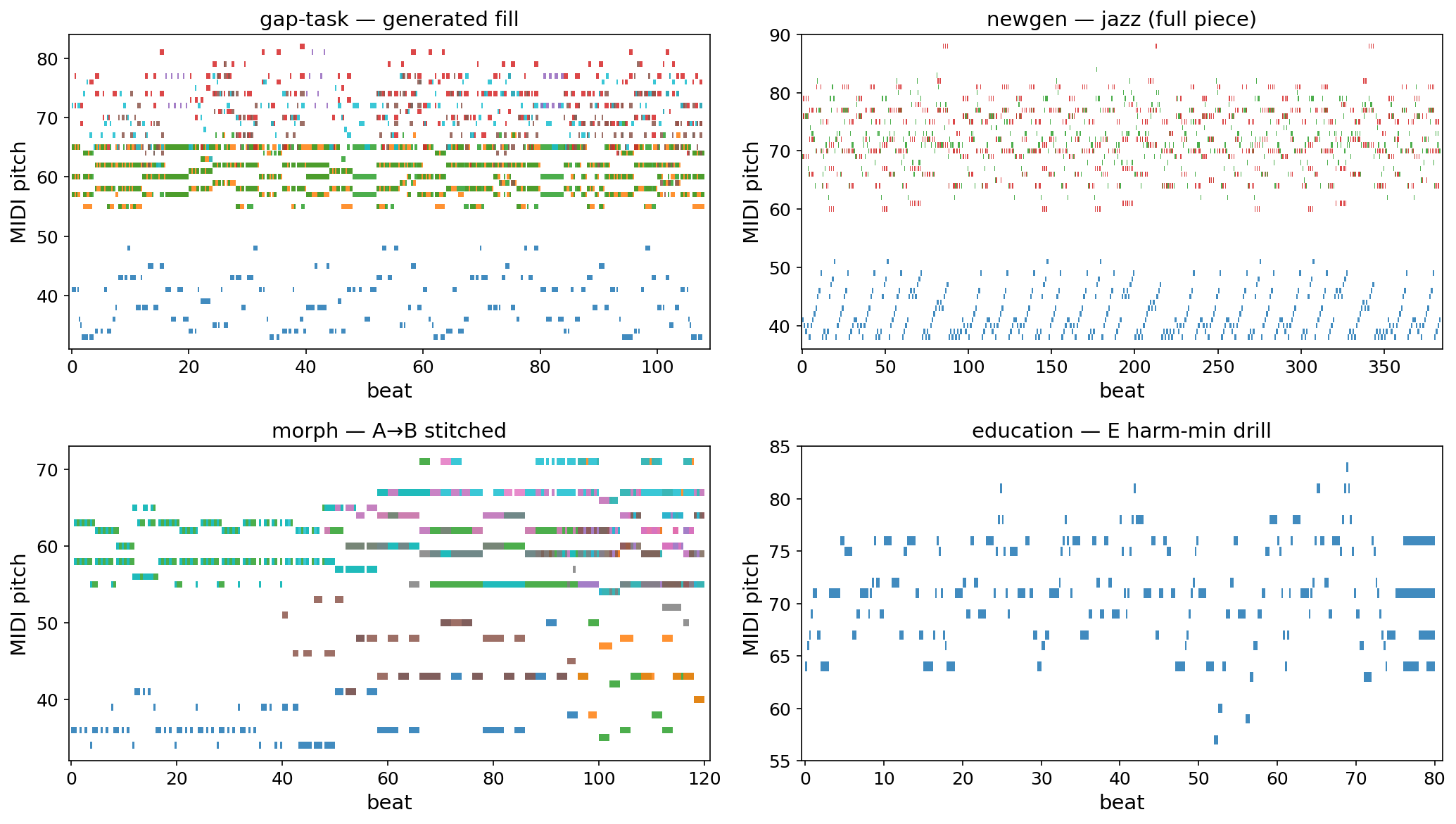}
    \caption{Representative generated outputs across four applications.}
    \label{fig:app_gallery}
\end{figure}

\begin{figure}[t]
    \centering
    \includegraphics[width=\linewidth]{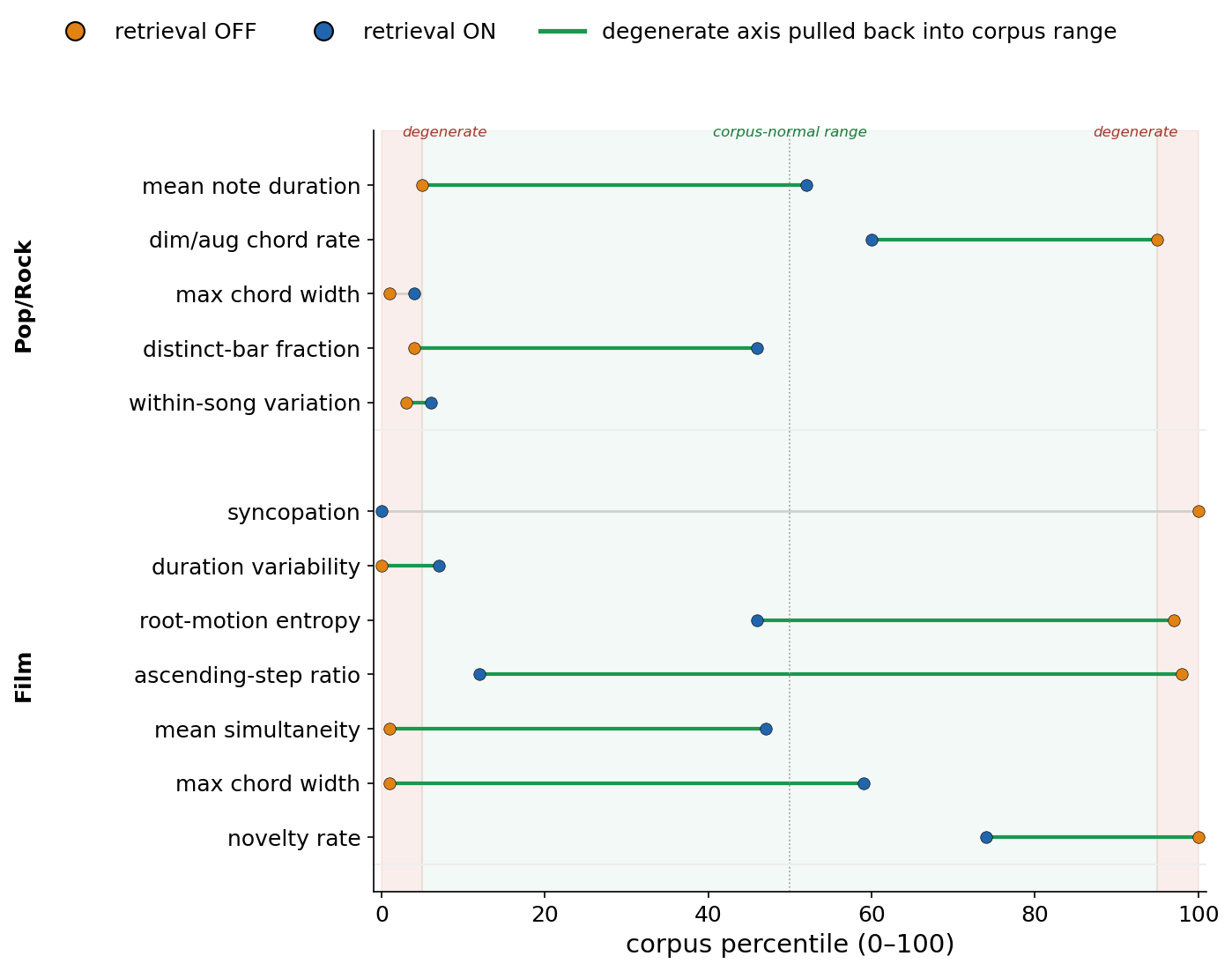}
    \caption{Retrieval de-degenerates full-piece generation.}
    \label{fig:retrieval_mechanism}
\end{figure}

\begin{figure}[t]
    \centering
    \includegraphics[width=\linewidth]{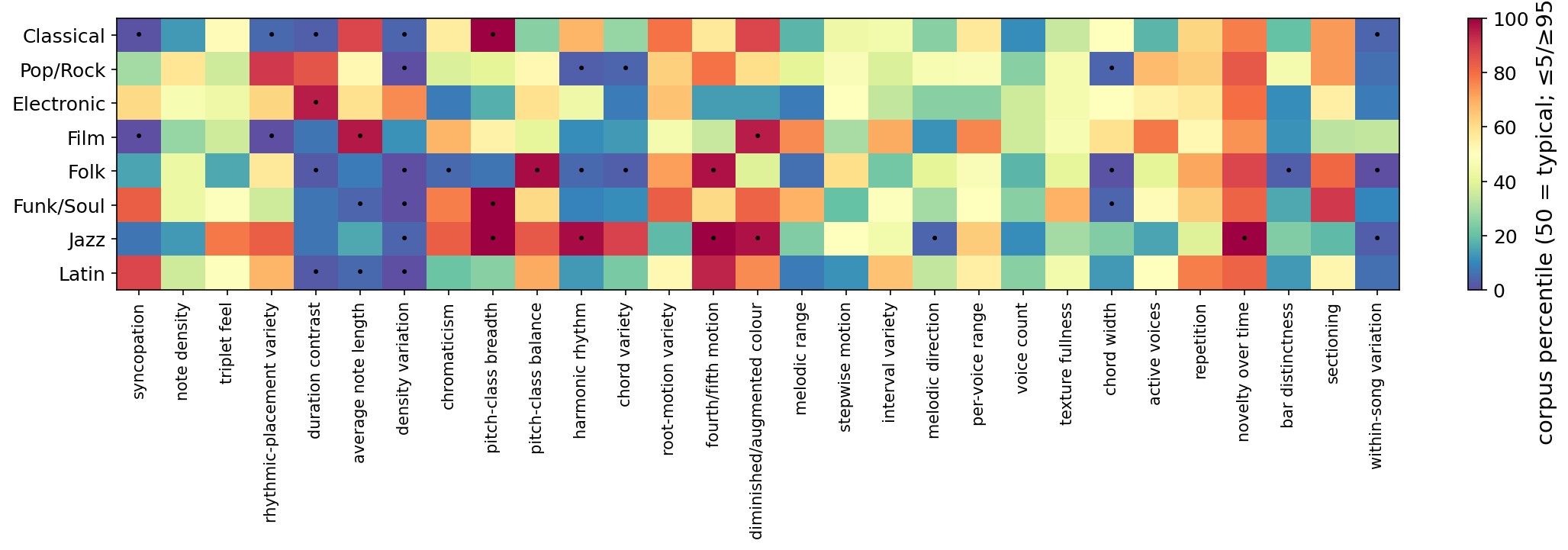}
    \caption{Per-piece axis profiles for generated outputs.}
    \label{fig:axis_profile_heatmap}
\end{figure}

\begin{figure}[t]
    \centering
    \includegraphics[width=\linewidth]{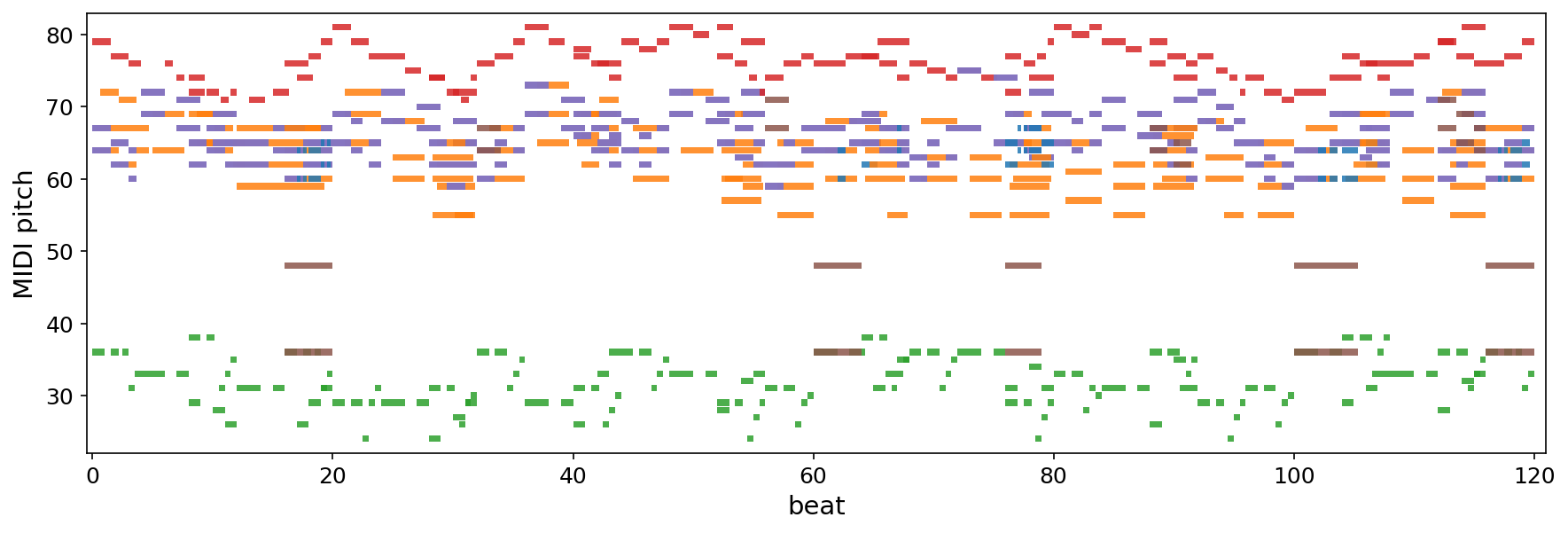}
    \caption{Interpretable failure case for a gap-fill continuation.}
    \label{fig:r2_roll}
\end{figure}

The appendix collects auxiliary results that complement the main text: \Cref{fig:genre_radar_overlaid} shows overlaid radar fingerprints across genres, \Cref{fig:app_gallery} summarizes representative outputs across the four applications, \Cref{fig:retrieval_mechanism} shows how retrieval de-degenerates full-piece generation, \Cref{fig:axis_profile_heatmap} visualizes per-piece axis profiles, and \Cref{fig:r2_roll} presents an interpretable gap-filling failure case. It also illustrates the Libretto grammar through three real generated outputs spanning single-voice pedagogy, multi-voice from-scratch generation, and anchored multi-voice gap filling. All examples are actual model outputs, trimmed for readability and edited only by elisions marked with \lit{...}.

\begin{grammarpanel}[label={box:education-drill}]{A. Education drill: single-voice pedagogy}

\panelnote{E harmonic minor, one \lit{Piano} voice. The excerpt shows the raised seventh \lit{D#}, the dominant \lit{B7}, and closing block chords written with \lit{+}.}

\begin{lstlisting}[style=librettoGrammar]
KEY: E minor | METER: 4/4 | TEMPO: 138 | GRID: 16th | BARS: 20
VOICES: Piano
@1 [Em]
  Piano: E4@1>1 F#4@2>1 G4@3>1 A4@4>1 B4@5>2 G4@7>2 E4@9>4 B4@13>4
@2 [Em]
  Piano: B4@1>2 E5@3>2 D#5@5>4 G4@9>2 A4@11>2 B4@13>4
@3 [Am]
  Piano: A4@1>1 B4@2>1 C5@3>1 E5@4>1 C5@5>2 A4@7>2 E5@9>4 C5@13>4
@4 [Em]
  Piano: G4@1>2 D#5@3>2 E5@5>4 B4@9>2 G4@11>2 E4@13>4
@5 [Em]
  Piano: E4@1>1 G4@2>1 B4@3>1 E5@4>1 D#5@5>1 B4@6>1 G4@7>1 F#4@8>1 E4@9>4 B4@13>4
@6 [Am]
  Piano: C5@1>2 A4@3>2 E5@5>2 C5@7>2 A4@9>4 E5@13>4
@7 [B7]
  Piano: B4@1>1 D#5@2>1 F#5@3>1 A5@4>1 F#5@5>1 D#5@6>1 B4@7>1 A4@8>1 D#5@9>4 B4@13>4
@8 [Em]
  Piano: E5@1>2 B4@3>2 G4@5>2 E4@7>2 F#4@9>2 G4@11>2 B4@13>4
...
@20 [Em]
  Piano: E4+B4+E5@1>8 G4+B4+E5@9>4 E4+G4+B4+E5@13>4
\end{lstlisting}

\panelread{In a 16th-note 4/4 grid, slot 1 is beat 1 and the beats fall on slots 1, 5, 9, and 13. A duration of \lit{>4} is a quarter note, while \lit{>1} is a sixteenth note.}

\end{grammarpanel}

\begin{grammarpanel}[label={box:new-generation}]{B. New generation: funk/soul multi-voice band}

\panelnote{A 96-bar funk/soul generation in G minor with five declared voices: \lit{BASS}, \lit{GTR}, \lit{KEYS}, \lit{HORNS}, and \lit{LEAD}. The excerpt shows bars 9--12, where four active voices interlock over a \lit{Gm7--Cm7--D7} vamp: off-beat guitar stabs, seventh-chord keyboard comping, a syncopated bass with pickups, and a lead line above.}

\begin{lstlisting}[style=librettoGrammar]
KEY: G minor | METER: 4/4 | TEMPO: 102 | GRID: 16th | BARS: 96
VOICES: BASS, GTR, KEYS, HORNS, LEAD
...
@9 [Gm7]
  BASS : G1@1>2 G1@4>1 Bb1@7>1 D2@9>2 D2@12>1 F1@15>2
  GTR  : Bb3+D4+F4@3>1 Bb3+D4+F4@7>1 Bb3+D4+F4@11>1 Bb3+D4+F4@15>1
  KEYS : G3+Bb3+D4+F4@3>2 G3+Bb3+D4+F4@11>2
  LEAD : D5@3>2 F5@7>1 G5@9>2 F5@13>1 D5@15>2
@10 [Cm7]
  BASS : C2@1>2 C2@4>1 Eb2@7>1 G2@9>2 G2@12>1 Bb1@15>2
  GTR  : Eb4+G4+Bb4@3>1 Eb4+G4+Bb4@7>1 Eb4+G4+Bb4@11>1 Eb4+G4+Bb4@15>1
  KEYS : C3+Eb3+G3+Bb3@3>2 C3+Eb3+G3+Bb3@11>2
  LEAD : G5@3>2 Eb5@7>1 D5@9>2 C5@13>3
@11 [Gm7]
  BASS : G1@1>2 G1@4>1 Bb1@7>1 D2@9>2 D2@12>1 F1@15>2
  GTR  : Bb3+D4+F4@3>1 Bb3+D4+F4@7>1 Bb3+D4+F4@11>1 Bb3+D4+F4@15>1
  KEYS : G3+Bb3+D4+F4@3>2 G3+Bb3+D4+F4@11>2
  LEAD : Bb4@3>2 D5@7>1 F5@9>2 G5@13>1 F5@15>2
@12 [D7]
  BASS : D2@1>2 D2@4>1 C#2@7>1 A1@9>2 A1@12>1 D2@15>2
  GTR  : F#3+A3+C4@3>1 F#3+A3+C4@7>1 F#3+A3+C4@11>1 F#3+A3+C4@15>1
  KEYS : D3+F#3+A3+C4@3>2 D3+F#3+A3+C4@11>2
  LEAD : A4@3>2 C5@7>1 D5@9>2 C#5@13>1 A4@15>2
...
\end{lstlisting}

\panelread{Slots 3, 7, 11, and 15 are off-beat sixteenth-grid positions between the main beats. The \lit{GTR} and \lit{KEYS} hits on those slots create the funk push, while \lit{+}-joined pitches encode chord voicings.}

\end{grammarpanel}

\begin{grammarpanel}[label={box:gap-fill}]{C. Gap filling: anchored multi-voice fill}

\panelnote{A generated region filled into an existing song's voice set. The excerpt shows a 9-voice adaptive-grid grammar with anonymized source tracks such as \lit{Part2} and \lit{Part6}, preserving the original multi-track structure while avoiding leakage from track names.}

\begin{lstlisting}[style=librettoGrammar]
KEY: Bb major | METER: 4/4 | TEMPO: 200 | GRID: 16th (adaptive) | BARS: 27
VOICES: Bass Gtr, Part2, Guitar 3, Guitar 2, Guitar 1, Part6, Part7, Strings, Marimba
@1 [F | Am]
  Bass Gtr: F2@1>3 F2@5>1 C2@6>1 B1@7>1 A1@8>4 A1@13>2 A1@15>1
  Part2   : A3+C4+F4@1>6 G3+A3+C4+E4@9>6
  Guitar 3: A3+C4+F4@1>2 A3+C4+F4@4>1 C4+F4@6>1 A3+C4+E4@9>2 C4+E4@11>1 A3+C4+E4@13>3
  Guitar 1: C5+F5@3>1 A4+C5@7>2 E5@13>3
  Part6   : C5@1>1 C5@5>2 D5@9>1 E5@11>1 G4@13>2 F4@16>1
@2 [Bb]
  Bass Gtr: Bb1@1>2 Bb1@5>1 F2@7>2 Bb1@9>3 C2@13>2 D2@15>1
  Part2   : Bb3+D4+F4@1>8 G3+Bb3+D4@9>5 A3+D4+F4@15>2
  Guitar 3: Bb3+D4+F4@1>1 D4+F4@3>1 Bb3+D4+F4@5>2 Bb3+D4+F4@8>1 G3+Bb3+D4@9>2 Bb3+D4+F4@11>1 Bb3+D4@13>2 F4@16>1
...
\end{lstlisting}

\panelread{The label \lit{[F | Am]} denotes a mid-bar harmonic change. Names such as \lit{Part2} and \lit{Part6} are anonymized real source tracks, used to keep the gap-filling task leakage-clean while preserving the original voice structure.}

\end{grammarpanel}

\begin{figure}[t]
    \centering
    \includegraphics[width=0.49\linewidth]{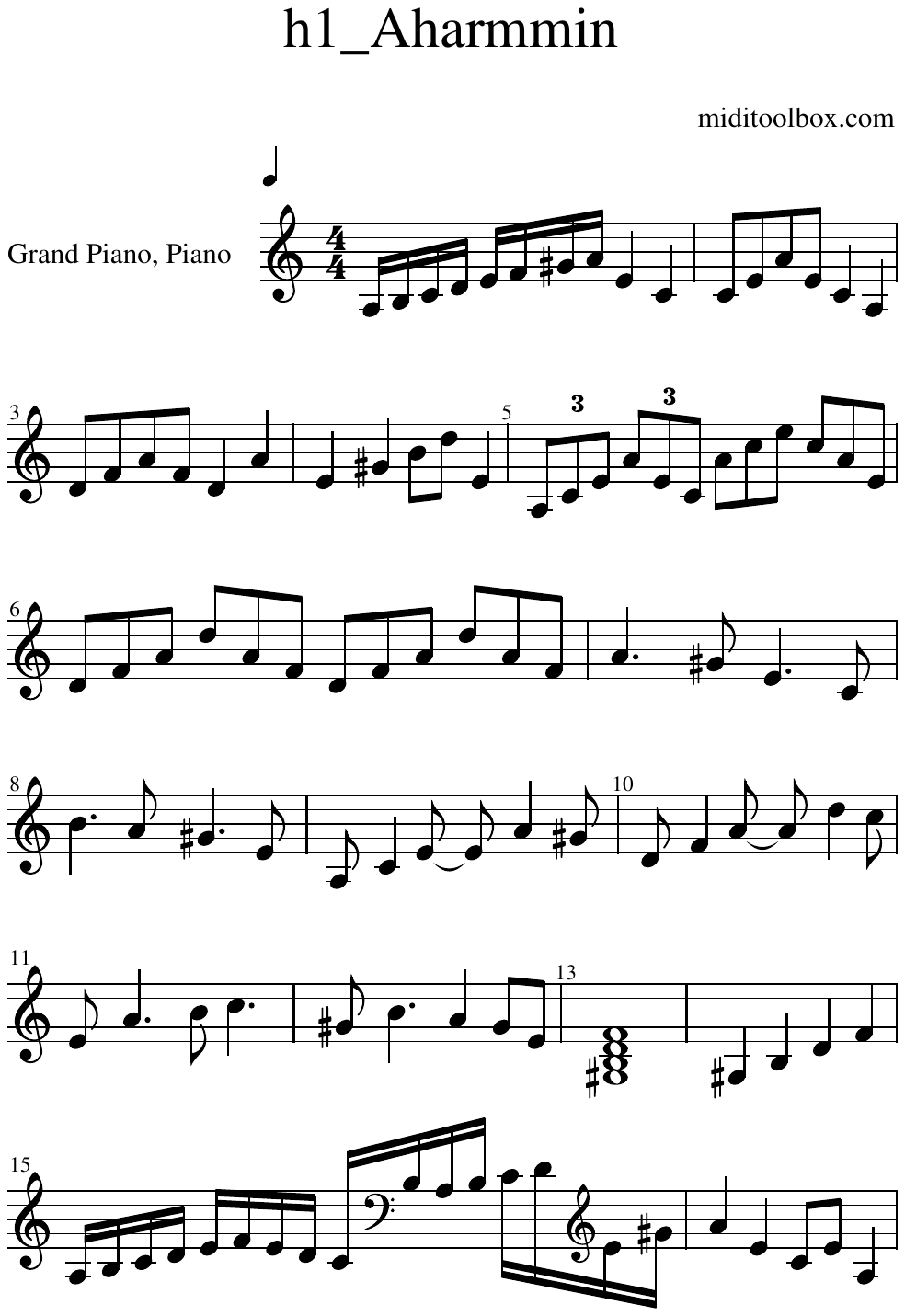}
    \hfill
    \includegraphics[width=0.49\linewidth]{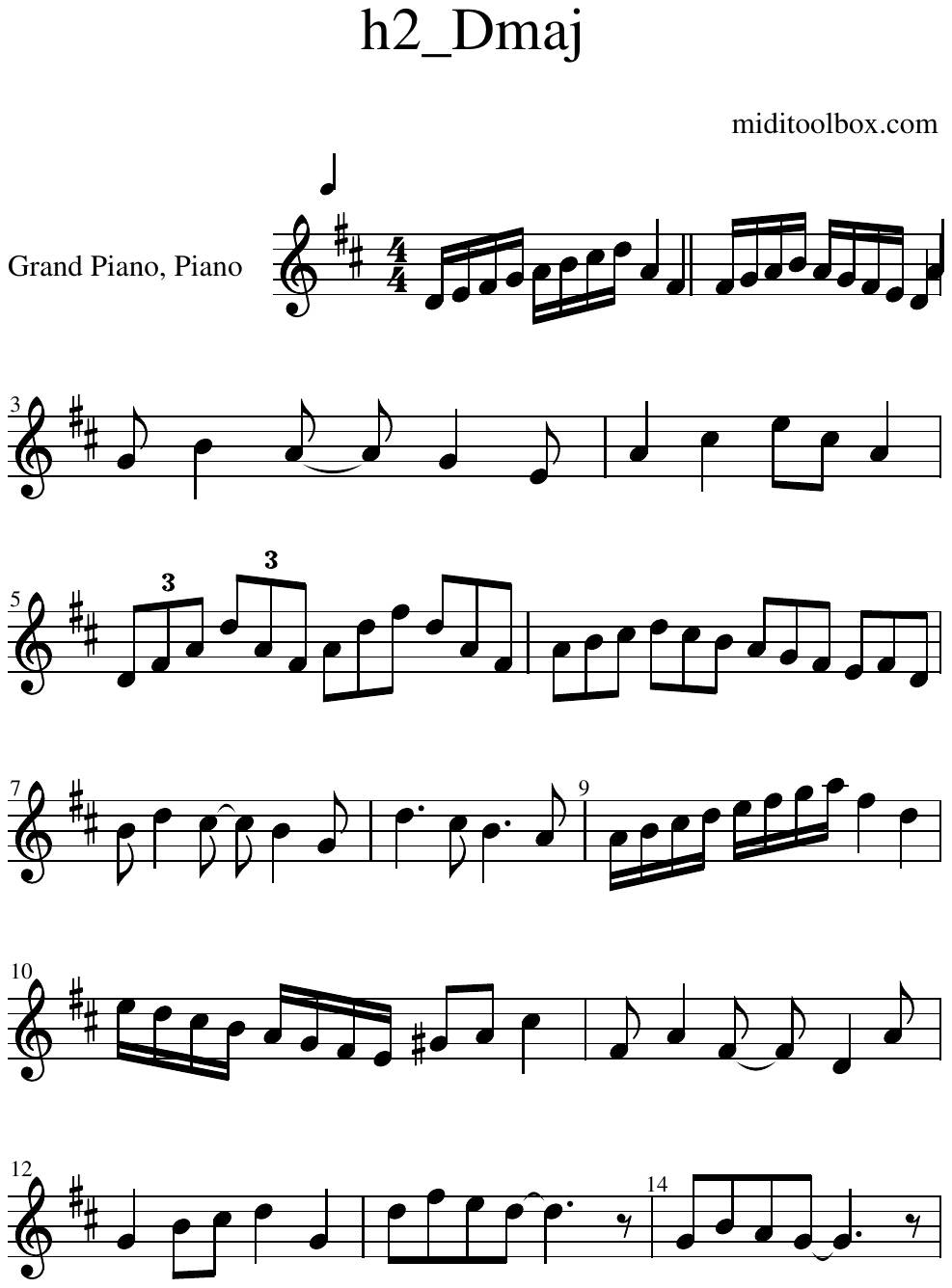}
    \caption{Educational music-generation examples for targeted harmonic concepts.}
    \label{fig:education_examples}
\end{figure}

\section{Metric Definitions}
\label{app:metrics}

This appendix defines the structural axes, percentile fingerprint, copy-risk score, and calibrated gates used in the experiments. All quantities are computed from the Libretto grammar tokens and are descriptive rather than aesthetic.

\paragraph{Notation.}
Let \(P=\{e\}\) be the set of parsed note events. Each event has bar \(b(e)\), within-bar onset \(o(e)\), absolute onset \(t(e)\), duration \(d(e)\), MIDI pitch \(m(e)\), pitch class \(pc(e)=m(e)\bmod 12\), and voice \(v(e)\). Let \(Q\) be beats per bar, \(\mathcal{B}\) the set of bars, \(N_b=\card{\mathcal{B}}\), \(V=\{v(e):e\in P\}\), and
\[
\mathcal{O}=\{(v(e),t(e)):e\in P\}, \qquad n_{\mathrm{on}}=\card{\mathcal{O}}.
\]
Let \(D=(d(e))_{e\in P}\). Means and standard deviations are population statistics. For a count vector \(c\), define normalized entropy
\[
H(c)=
\begin{cases}
-\dfrac{\sum_{i:c_i>0}p_i\log_2 p_i}{\log_2 k'}, & k'>1,\\
0, & k'\leq 1,
\end{cases}
\qquad
p_i=\frac{c_i}{\sum_j c_j}, \quad k'=\card{\{i:c_i>0\}}.
\]
For event set \(E\), define duration-weighted pitch-class mass \(w_E(p)=\sum_{e\in E:pc(e)=p}d(e)\), and
\[
\Prom(w)=\{p:w(p)\geq 0.30\max_q w(q)\}.
\]
For each voice \(u\), let \(\mu(u)=\Mean\{m(e):v(e)=u\}\) and \(\chi(u)=\card{\{e:v(e)=u\}}/\card{\{t(e):v(e)=u\}}\). The bass is the lowest-\(\mu\) voice. The melody is the highest-\(\mu\) voice among voices with \(\chi<1.4\) and at least 8 distinct onsets, or otherwise the highest-\(\mu\) voice. Let \(s_1,\ldots,s_L\) be the melody line, taking the highest MIDI pitch at each melody-voice onset.

For each bar \(b\), define the note set
\[
A_b=\{(v(e),o(e),m(e)):b(e)=b\},
\]
and the self-similarity matrix
\[
S_{ij}=\frac{\card{A_i\cap A_j}}{\card{A_i\cup A_j}}.
\]

\paragraph{Rhythm axes.}
\[
\begin{aligned}
\textsc{Syncopation Rate} &= \frac{\card{\{(u,t)\in\mathcal{O}:t-\lfloor t\rfloor\neq 0\}}}{n_{\mathrm{on}}},\\
\textsc{Onset Density} &= \frac{n_{\mathrm{on}}}{N_b},\\
\textsc{Triplet Share} &= \frac{n_{\mathrm{trip}}}{n_{\mathrm{trip}}+n_{\mathrm{bin}}},\\
\textsc{Onset Position Entropy} &= H\!\left(\Hist\!\left(\Round((t(e)\bmod Q)/0.25)\right)\right),\\
\textsc{Duration CV} &= \frac{\Std(D)}{\Mean(D)},\\
\textsc{Mean Duration} &= \Mean(D),\\
\textsc{Density Variability} &= \frac{\Std((n_b)_b)}{\Mean((n_b)_b)}, \qquad n_b=\card{\{e:b(e)=b\}}.
\end{aligned}
\]

\paragraph{Harmony axes.}
Let \(w(p)=w_P(p)\), \(W=\sum_p w(p)\), and \(S_{\mathrm{maj}}=\{0,2,4,5,7,9,11\}\). Split the piece into half-bars \(h_k\) and define \(C_k=\Prom(w_{h_k})\). Let \(r_b\) be the lowest bass MIDI pitch in bar \(b\), reduced modulo 12, and let \(\tau_b=(r_{b+1}-r_b)\bmod 12\).
\[
\begin{aligned}
\textsc{Chromaticism}
&=1-\frac{\max_{r\in\{0,\ldots,11\}}\sum_{i\in S_{\mathrm{maj}}}w((r+i)\bmod 12)}{W},\\
\textsc{Distinct Pitch Classes} &= \card{\{p:w(p)>0\}},\\
\textsc{Pitch-Class Entropy} &= H((w(0),\ldots,w(11))),\\
\textsc{Chord Change Rate} &=
\frac{\card{\{k:C_k\neq C_{k+1},\,C_k\neq\emptyset,\,C_{k+1}\neq\emptyset\}}}{2N_b-1},\\
\textsc{Chord Vocabulary Density} &= \frac{\card{\{C_k:C_k\neq\emptyset\}}}{N_b},\\
\textsc{Root-Motion Entropy} &= H(\Hist((\tau_b)_b)),\\
\textsc{Fourth-Motion Rate} &= \frac{\card{\{b:\tau_b=5\}}}{\card{\{\tau_b\}}}.
\end{aligned}
\]
The diminished/augmented color axis is
\[
\begin{aligned}
\textsc{Diminished-Augmented Color}
&=
\frac{D_{\mathrm{dim}}+\min\{D_{\mathrm{aug}},N_b\}}{N_b},\\
D_{\mathrm{dim}}
&=
\sum_b
\indic{\exists r:\{r,r+3,r+6\}\subseteq \Prom(w_b)},\\
D_{\mathrm{aug}}
&=
\sum_b
\card{\{r:\{r,r+4,r+8\}\subseteq \Prom(w_b)\}}.
\end{aligned}
\]
with pitch classes interpreted modulo 12.

\paragraph{Melody axes.}
Let \(\iota_j=s_{j+1}-s_j\) and \(M=\{j:\iota_j\neq 0\}\).
\[
\begin{aligned}
\textsc{Pitch Range} &= \max_{e\in P}m(e)-\min_{e\in P}m(e),\\
\textsc{Step Ratio} &= \frac{\card{\{j\in M:\card{\iota_j}\leq 2\}}}{\card{M}},\\
\textsc{Interval Entropy} &= H\!\left(\Hist((\min\{\card{\iota_j},12\})_j)\right),\\
\textsc{Ascending Ratio} &= \frac{\card{\{j\in M:\iota_j>0\}}}{\card{M}},\\
\textsc{Melody-Voice Range} &= \max_{e:v(e)=\mathrm{melody}}m(e)-\min_{e:v(e)=\mathrm{melody}}m(e).
\end{aligned}
\]
If \(M=\emptyset\), \(\textsc{Ascending Ratio}=0.5\).

\paragraph{Texture axes.}
\[
\begin{aligned}
\textsc{Voice Count} &= \card{V},\\
\textsc{Mean Simultaneity} &= \frac{\card{P}}{n_{\mathrm{on}}},\\
\textsc{Maximum Chord Width} &=
\max_{(u,t):\,\card{P_{u,t}}\geq 2}
\left(\max_{e\in P_{u,t}}m(e)-\min_{e\in P_{u,t}}m(e)\right),\\
\textsc{Active Voice Density} &= \Mean_{b\in\mathcal{B}}\card{\{v(e):b(e)=b\}},
\end{aligned}
\]
where \(P_{u,t}=\{e:v(e)=u,t(e)=t\}\).

\paragraph{Form axes.}
\[
\begin{aligned}
\textsc{Self-Similarity} &= \Mean_{i<j}S_{ij},\\
\textsc{Novelty Rate} &= \Mean_i(1-S_{i,i+1}),\\
\textsc{Distinct-Bar Fraction} &= \frac{\card{\{A_b:b\in\mathcal{B}\}}}{N_b}.
\end{aligned}
\]
For section density, let \(L=\min\{4,\lfloor N_b/4\rfloor\}\). Define a checkerboard novelty curve
\[
\mathrm{nov}(c)=
\frac{1}{\card{\mathrm{cells}}}
\sum_{\alpha,\beta\in[-L,L)}
\mathrm{sgn}(\alpha,\beta)S_{c+\alpha,c+\beta},
\]
where \(\mathrm{sgn}(\alpha,\beta)=+1\) if \((\alpha<0)=(\beta<0)\), and \(-1\) otherwise. A peak is a local maximum with \(\mathrm{nov}(c)\geq \Mean(\mathrm{nov})+0.5\Std(\mathrm{nov})\). Then
\[
\textsc{Sections per 100 Bars}=\frac{\#\mathrm{peaks}+1}{N_b}\cdot 100.
\]

\paragraph{Within-song variation.}
Split the piece into \(W\) equal-bar windows. For each window \(w\), compute a base vector \(x_w\) over the rhythm, harmony, melody, texture, and form axes. For each base axis \(a\), let \(\sigma_a=\Std_w(x_w[a])\), and let \(SD_a\) be the corpus standard deviation of axis \(a\). The within-song variation axis is
\[
\textsc{Within-Song Variation}=\Mean_a\frac{\sigma_a}{SD_a}.
\]

\paragraph{Percentile fingerprint.}
For axis \(a\), let \(v_a=\mathrm{axis}_a(P)\), and let \(\mathrm{col}_a\) be the frozen 314-song corpus column for that axis. The percentile coordinate is
\[
\mathrm{pct}_a(P)
=
\Round\left(
100\cdot
\frac{1}{314}
\card{\{x\in\mathrm{col}_a:x\leq v_a\}}
\right).
\]
The fingerprint is \((\mathrm{pct}_a(P))_{a=1}^{29}\). An axis is a degenerate extreme iff \(\mathrm{pct}_a(P)\leq 5\) or \(\mathrm{pct}_a(P)\geq 95\).

\paragraph{Copy risk and gates.}
Represent a piece by \(g[b]=\{(\Round(o(e),2),m(e)):b(e)=b\}\). For real song \(S\),
\[
A(g,g_S,\delta)=
\frac{\sum_b\card{g[b]\cap g_S[b+\delta]}}{\card{P}},
\qquad
\mathrm{slide}(g,g_S)=\max_\delta A(g,g_S,\delta).
\]
The copy-risk score is
\[
\mathrm{copy\ risk}(P)
=
\max\left\{
\max_{S\in\mathrm{cited}}\mathrm{slide}(g,g_S),
\max_{S\in\mathrm{top25}}\mathrm{slide}(g,g_S),
\mathrm{slide}(g,g_{\mathrm{ref}})
\right\}.
\]
For genre \(g\), gates are calibrated from real songs:
\[
C1_g=\clipthree{\lceil\qntl{0.85}{\mathrm{extreme\ counts}(g)}\rceil}{3}{6},
\]
\[
F_g=\clipthree{\lfloor\qntl{0.15}{\mathrm{band\ occupancy}(g)}\rfloor}{3}{6},
\]
\[
T_g=\clipthree{1.20\cdot\qntl{0.90}{\mathrm{copy\ risk}(g)}}{0.30}{0.45}.
\]
A generated piece passes the shared gates when \(n_{\mathrm{extreme}}(P)\leq C1_g\), \(\mathrm{fit}(P,g)\geq F_g\), and \(\mathrm{copy\ risk}(P)<T_g\).

\end{document}